# Chapter XX

# Gaussian Beam Transmission through a Gyrotropic-Nihility Finely-Stratified Structure

**Vladimir R. Tuz**[1,2], **Volodymyr I. Fesenko**[1,3]

[1]Institute of Radio Astronomy of NASU, Kharkiv, Ukraine

[2]School of Radio Physics, Karazin Kharkiv National University, Kharkiv, Ukraine,
tvr@rian.kharkov.ua

[3]Lab. "Photonics", Kharkiv National University of Radio Electronics, Kharkiv, Ukraine,
fesenko@kture.kharkov.ua; fesenko_vladimir@mail.ru

**Abstract** The three-dimensional Gaussian beam transmission through a ferrite-semiconductor finely-stratified structure being under an action of an external static magnetic field in the Faraday geometry is considered. The beam field is represented by an angular continuous spectrum of plane waves. In the long-wavelength limit, the studied structure is described as a gyroelectromagnetic medium defined by the effective permittivity and effective permeability tensors. The investigations are carried out in the frequency band where the real parts of the on-diagonal elements of both effective permittivity and effective permeability tensors are close to zero while the off-diagonal ones are non-zero. In this frequency band the studied structure is referred to a gyrotropic-nihility medium. It is found out that a Gaussian beam keeps its parameters unchanged (beam width and shape) when passing through the layer of such a medium except of a portion of the absorbed energy.

## 1.1 Introduction

The conception of *nihility* was firstly introduced in the paper [1] for a hypothetical medium, in which the following constitutive relations hold $\vec{D} = 0$, $\vec{B} = 0$. So, nihility is the electromagnetic nilpotent, and the wave propagation cannot occur in nihility, because $\nabla \times \vec{E} = 0$ and $\nabla \times \vec{H} = 0$ in the absence of sources therein.

Further, in [2], this conception of nihility was extended for an isotropic chiral medium whose constitutive relations are: $\vec{D} = \varepsilon\vec{E} + i\rho\vec{H}$, $\vec{B} = \mu\vec{H} - i\rho\vec{E}$, where $\rho$ is the chirality parameter. Thus, a possible way for composing such a medium in the microwave band was proposed using canonical chiral wire particles. The effective material parameters are calculated on the basis of the Maxwell-Garnett mixing rule, and in a certain narrow frequency band it is found out that the real parts of both effective permittivity and effective permeability become close to zero ($\varepsilon' \approx 0$, $\mu' \approx 0$) while the real part of the chirality parameter is maintained at a finite value ($\rho' \neq 0$). It was revealed that in such an isotropic *chiral-nihility* medium there are two eigenwaves with right (RCP) and left (LCP) circularly polarized states, whose propagation constants depend only on the chirality parameter, and these propagation constants of the RCP ($\gamma^+$) and LCP ($\gamma^-$) waves are equal in magnitude but opposite in sign to each other ($\gamma^\pm = \pm k_0 \rho = \pm \gamma$). Thereby one of these eigenwaves experiences the forward propagation while the other one experiences the backward propagation. Here, the sign of the chirality parameter, which in turn depends on the chiral particles handedness, determines which of the eigenwaves appears as a backward propagating one. In particular, this feature results in some exotic characteristics in the wave transmission through and reflection from a single layer and multilayer systems which consist of such a chiral-nihility medium [3, 4].

Besides chiral media, the circularly polarized eigenwaves are also inherent to magneto-optic gyrotropic materials (e.g. ferrites or semiconductors) in the presence of an external static magnetic field, when this field is biased to the specimen in the longitudinal geometry relative to the direction of wave propagation (in the Faraday configuration) [5]. Such gyrotropic media are characterized by the permeability or permittivity tensor $\vec{D} = \hat{\varepsilon}\vec{E}$, $\vec{B} = \hat{\mu}\vec{H}$ with non-zero off-diagonal elements (gyrotropic parameters). Apart from getting double-negative conditions [6-10], combining together gyromagnetic (ferrite) and gyroelectric (semiconductor) materials into a certain unified gyroelectromagnetic structure [6] allows one to reach the *gyrotropic-nihility* effect within a narrow frequency band [11]. In particular, in a finely stratified ferrite-semiconductor structure such a condition is valid in the microwave band nearly the frequencies of ferromagnetic and plasma resonances. In this case the real parts of on-diagonal elements of both effective permeability and effective permittivity tensors of such an artificial medium simultaneously acquire zero while the off-diagonal ones are non-zero. It is revealed that in this medium the backward propagation can appear for one of the circularly polarized eigenwaves which leads to some unusual optical features of the system and provides an enhancement of the polarization rotation, impedance matching to free space, and complete light transmission.

Since a gyrotropic-nihility medium with appropriate parameters can support backward propagating eigenwaves and is impedance matched to free space, it becomes substantial to study the focusing properties of a finite thickness slab in which the gyrotropic-nihility condition holds [12]. It involves consideration of the field in the form of a spatially finite wave beam, in particular, as a Gaussian beam which is presented as a continual superposition of plane waves. On the other hand, it is also known that there are several beam phenomena such as displacement of the beam axis, beam splitting, focal and angular shifts which are not found in the reflection and transmission of separate plane waves [13-16], and so they require particular consideration. These studies are usually based on a two-dimensional beam formulation, which is quite efficient [17, 18]. Nevertheless, in gyrotropic media a three-dimensional model of beam representation should be considered to take into account the polarization effects and to predict the change in the ellipticity of the scattered beam [19, 20].

In this chapter, we demonstrate the phenomenon of the three-dimensional Gaussian beam transmission through a ferrite-semiconductor finely-stratified structure being under an action of an external static magnetic field biased along the structure periodicity. The investigations are carried out for two different frequencies. The first one is chosen to be far from frequencies of the ferromagnetic and plasma resonances and the second one is selected to be at the gyrotropic-nihility frequency. The main goal is to show that such a finely-stratified structure is able to tunnel a Gaussian beam practically without any distortion of its form when the gyrotropic-nihility condition holds.

## 1.2 Problem Formulation and Methods of Solution

### *1.2.1 Magnetic Multilayer Structure under Study*

A stack of $N$ identical double-layer slabs (unit cells) which are arranged periodically along the $z$ axis is investigated (Fig. 1). Each unit cell is composed of ferrite (with constitutive parameters $\varepsilon_1$, $\hat{\mu}_1$) and semiconductor (with constitutive parameters $\hat{\varepsilon}_2$, $\mu_2$) layers with thicknesses $d_1$ and $d_2$, respectively. The structure's period is $L = d_1 + d_2$, and in the $x$ and $y$ directions the system is infinite. We suppose that the structure is finely-stratified, i.e. its characteristic dimensions $d_1$, $d_2$ and $L$ are significantly smaller than the wavelength in the corresponding layer $d_1 \ll \lambda$, $d_2 \ll \lambda$, $L \ll \lambda$ (the long-wavelength limit). An external static magnetic field $\vec{M}$ is directed along the $z$-axis. The input $z \leq 0$ and output $z \geq NL$ half-spaces are homogeneous, isotropic and have constitutive parameters $\varepsilon_0$, $\mu_0$.

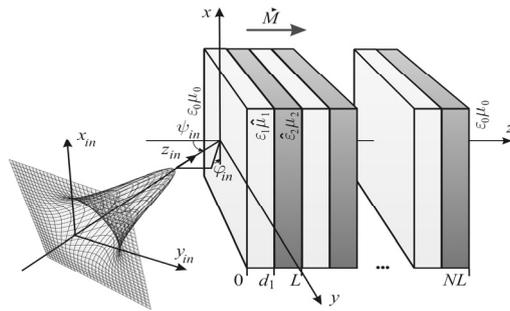

**Fig. 1.** A periodic stack of one-dimensional double-layer ferrite-semiconductor structure under the Gaussian beam illumination.

We use common expressions for constitutive parameters of normally magnetized ferrite and semiconductor layers with taking into account the losses. They are defined in the form [21-23]:

$$\varepsilon_1 = \varepsilon_f, \qquad \hat{\mu}_1 = \begin{pmatrix} \mu_1^T & -i\alpha & 0 \\ i\alpha & \mu_1^T & 0 \\ 0 & 0 & \mu_1^L \end{pmatrix}, \qquad \hat{\varepsilon}_2 = \begin{pmatrix} \varepsilon_2^T & -i\beta & 0 \\ i\beta & \varepsilon_2^T & 0 \\ 0 & 0 & \varepsilon_2^L \end{pmatrix}, \qquad \mu_2 = \mu_s, \qquad (1)$$

where for ferrite the auxiliary values are $\mu_1^T = 1 + \chi' + i\chi''$, $\chi' = \omega_0 \omega_m \left[ \omega_0^2 - \omega^2(1-b^2) \right] D^{-1}$, $\chi'' = \omega \omega_m b \left[ \omega_0^2 - \omega^2(1+b^2) \right] D^{-1}$, $\alpha = \Omega' + i\Omega''$, $\Omega' = \omega \omega_m \left[ \omega_0^2 - \omega^2(1+b^2) \right] D^{-1}$, $\Omega'' = 2\omega^2 \omega_0 \omega_m b D^{-1}$, $D = \left[ \omega_0^2 - \omega^2(1+b^2) \right]^2 + 4\omega_0^2 \omega^2 b^2$, $\omega_0$ is the Larmor frequency and $b$ is a dimensionless damping constant; for semiconductor layers the auxiliary values are $\varepsilon_2^L = \varepsilon_0 \left[ 1 - \omega_p^2 [\omega(\omega+i\nu)]^{-1} \right]$, $\beta = \varepsilon_0 \omega_p^2 \omega_c \left[ \omega((\omega+i\nu)^2 - \omega_c^2) \right]^{-1}$, $\varepsilon_2^T = \varepsilon_0 \left[ 1 - \omega_p^2 (\omega+i\nu) [\omega((\omega+i\nu)^2 - \omega_c^2)]^{-1} \right]$, $\varepsilon_0$ is the part of permittivity attributed to the lattice, $\omega_p$ is the plasma frequency, $\omega_c$ is the cyclotron frequency and $\nu$ is the electron collision frequency in plasma.

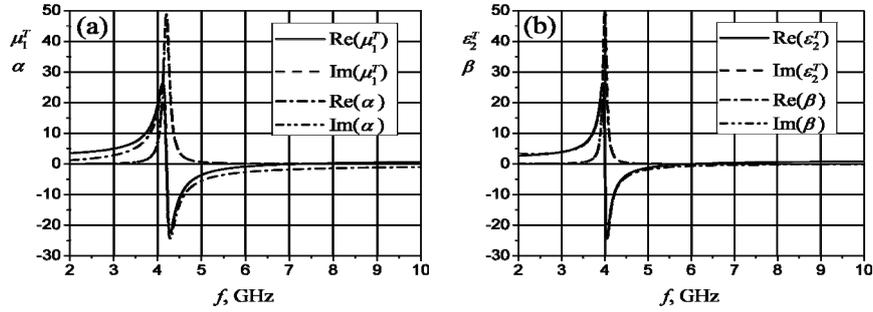

**Fig. 2.** Frequency dependences of the permeability and permittivity of ferrite (**a**) and semiconductor (**b**) layers, respectively. We use typical parameters for these materials in the microwave region. For the ferrite layers, under saturation magnetization of 2000 G, parameters are $\omega_0/2\pi = 4.2$ GHz, $\omega_m/2\pi = 8.2$ GHz, $b = 0.02$, $\varepsilon_f = 5.5$. For the semiconductor layers, parameters are: $\omega_p/2\pi = 4.5$ GHz, $\omega_c/2\pi = 4.0$ GHz, $\nu/2\pi = 0.05$ GHz, $\varepsilon_0 = 1.0$, $\mu_s = 1.0$.

The frequency dependences of the permeability and permittivity parameters calculated using Eq. (1) are presented in Fig. 2. Note that the values of $\text{Im}(\mu_1^T)$, $\text{Im}(\alpha)$ and $\text{Im}(\varepsilon_2^T)$, $\text{Im}(\beta)$ are so close to each other that the curves of their frequency dependences coincide in the corresponding figures.

## *1.2.2 Gaussian Beam Representation*

The auxiliary coordinate system $x_{in}$, $y_{in}$, $z_{in}$ (see, Fig. 1) is introduced to describe the incident beam field [13, 14, 20]. In it, the incident field $\vec{\psi}_{in} = \vec{E}_{in}, \vec{H}_{in}$ is defined as a continued sum of the partial plane waves with the spectral parameter $\vec{\kappa}_{in}$ (it has a sense of the transverse wave vector of the partial plane wave):

$$\vec{\psi}_{in} = \vec{\upsilon} \iint_{-\infty}^{\infty} U(\vec{\kappa}_{in}) \exp\left[i\vec{\kappa}_{in}(\vec{r}_{in} + \vec{a}_{in}) + i\gamma_{in}(z_{in} + a_3)\right] d\vec{\kappa}_{in}. \qquad (2)$$

In Eq. (2) the vector $\vec{\upsilon}$ is related to $E$ ($\vec{\upsilon} = \vec{e}_{in}$) or $H$ ($\vec{\upsilon} = \vec{h}_{in}$) field, respectively; $\vec{e}_{in} = \vec{P}V_p - \vec{b}_{in} \times \vec{P}V_s$, $\vec{h}_{in} = \vec{P}V_s + \vec{b}_{in} \times \vec{P}V_p$ where the vector $\vec{P} = \vec{z}_0 \times \vec{n}$ describes the field polarization. In the structure's coordinates $x$, $y$, $z$, the vector $\vec{n}$ is characterized via the next components $(\cos\theta_{in}\cos\varphi_{in}, \cos\theta_{in}\sin\varphi_{in}, 0)$, where $\theta_{in} = 90° - \psi_{in}$; $\vec{z}_0$ is the basis vector of $z$-axis, and the vector $\vec{b}_{in} = (\cos\theta_{in}\cos\varphi_{in}, \cos\theta_{in}\sin\varphi_{in}, -\sqrt{\varepsilon_0\mu_0 - \cos^2\theta_{in}})$ describes the direction of the incident beam propagation; $U(\vec{\kappa}_{in})$ is the spectral density of the beam in the plane $z_{in} = 0$; $\gamma_{in} = \sqrt{k_0^2 - \vec{\kappa}_{in} \cdot \vec{\kappa}_{in}}$, $0 < \arg(\sqrt{k_0^2 - \vec{\kappa}_{in} \cdot \vec{\kappa}_{in}}) < \pi$, and $\vec{a}_{in} = (a_1, a_2)$.

The transformation from the structure's coordinate system $\{x, y, z\}$ to the beam's one $\{x_{in}, y_{in}, z_{in}\}$ can be realized in the standard means [24]: by rotating around the $z$-axis on the angle $\varphi_{in}$; by rotating around the $y$-axis on the angle $\psi_{in}$; by shifting the point of origin to the point $(a_1, a_2, a_3)$. Taking into account the coordinate system transformation, the reflected and transmitted fields are obtained as follows:

$$\begin{aligned}\vec{\psi}_{ref} &= \vec{P}V_{\upsilon} \iint_{-\infty}^{\infty} U(\vec{\kappa}_{in}) R^{\upsilon\upsilon} \exp\left[i\vec{\kappa}\cdot\vec{r} - i\gamma z\right] d\vec{\kappa}_{in} \pm \\ &\quad \pm \vec{b}_{ref} \times \vec{P}V_{\upsilon'} \iint_{-\infty}^{\infty} U(\vec{\kappa}_{in}) R^{\upsilon'\upsilon} \exp\left[i\vec{\kappa}\cdot\vec{r} - i\gamma z\right] d\vec{\kappa}_{in}, \\ \vec{\psi}_{tr} &= \vec{P}V_{\upsilon} \iint_{-\infty}^{\infty} U(\vec{\kappa}_{in}) T^{\upsilon\upsilon} \exp\left[i\vec{\kappa}\cdot\vec{r} + i\gamma(z - NL)\right] d\vec{\kappa}_{in} \pm \\ &\quad \pm \vec{b}_{tr} \times \vec{P}V_{\upsilon'} \iint_{-\infty}^{\infty} U(\vec{\kappa}_{in}) T^{\upsilon'\upsilon} \exp\left[i\vec{\kappa}\cdot\vec{r} + i\gamma(z - NL)\right] d\vec{\kappa}_{in},\end{aligned} \qquad (3)$$

where $\gamma = \sqrt{k_0^2 - \vec{\kappa}\cdot\vec{\kappa}}$, $0 < \arg(\sqrt{k_0^2 - \vec{\kappa}\cdot\vec{\kappa}}) < \pi$; $R^{\upsilon\upsilon}$, $R^{\upsilon'\upsilon}$ and $T^{\upsilon\upsilon}$, $T^{\upsilon'\upsilon}$ are the complex reflection and transmission coefficients ($\upsilon, \upsilon' = s, p$) of the partial plane electromagnetic waves, respectively. They depend on the frequency of the incident field, angles $\psi_{in}$, $\varphi_{in}$ and other electromagnetic and geometric parame-ters of the structure. The coefficients with coincident indexes ($\upsilon\upsilon$) describe the transformation of the incident wave of the perpendicular ($\upsilon = s$) or the parallel ($\upsilon = p$) polarization into the co-polarized wave, and the coefficients with distinct indexes ($\upsilon'\upsilon$) describe the transformation of the incident wave into the cross-polarized wave at the structure output. The left and right indexes correspond to the polarization states of the incident and reflected (transmitted) waves, respectively. The corresponding reflection and transmission coefficients are determined through the rigorous solution of the Cauchy problem related to the tangential field components on the structure's boundaries; the reader is referred here to Refs. [5, 11] for further details.

### *1.2.3 Effective Medium Theory*

In the long-wavelength limit, when the characteristic dimensions of the structure ($d_1$, $d_2$, $L$) are *significantly smaller* than the wavelength in the corresponding layer ($d_1 \ll \lambda$, $d_2 \ll \lambda$, $L \ll \lambda$), the interactions of electromagnetic waves with a periodic gyromagnetic-gyroelectric structure can be described analytically using the effective medium theory. From the viewpoint of this theory, the periodic structure is represented approximately as an anisotropic (gyroelectromagnetic) *uniform* medium whose optical axis is directed along the structure periodicity, and this medium is described with some effective permittivity and permeability tensors $\hat{\varepsilon}$ and $\hat{\mu}$ [11]. By this means, the investigation of the wave interaction with an inhomogeneous periodic structure is reduced to the solution of the boundary-value problem of conjugations of an equivalent homogeneous anisotropic layer with surrounding spaces.

Let us consider a unit cell of the studied structure. It is made of two layers $0 \leq z \leq d_1$ and $d_1 \leq z \leq L$ of dissimilar materials whose constitutive relations are as follows:

$$\left.\begin{array}{l}\vec{D} = \varepsilon_1 \vec{E} \\ \vec{B} = \hat{\mu}_1 \vec{H}\end{array}\right\} \ 0 \leq z \leq d_1, \quad \left.\begin{array}{l}\vec{D} = \hat{\varepsilon}_2 \vec{E} \\ \vec{B} = \mu_2 \vec{H}\end{array}\right\} \ d_1 \leq z \leq L. \tag{4}$$

In general form, in the Cartesian coordinates, the system of Maxwell's equations for each layer has a form

$$\begin{aligned} ik_y H_z - \partial_z H_y &= -ik_0 (\hat{\varepsilon}_j \vec{E})_x, & ik_y E_z - \partial_z E_y &= ik_0 (\hat{\mu}_j \vec{H})_x, \\ \partial_z H_x - ik_x H_z &= -ik_0 (\hat{\varepsilon}_j \vec{E})_y, & \partial_z E_x - ik_x E_z &= ik_0 (\hat{\mu}_j \vec{H})_y, \\ ik_x H_y - ik_y H_x &= -ik_0 (\hat{\varepsilon}_j \vec{E})_z, & ik_x E_y - ik_y E_x &= ik_0 (\hat{\mu}_j \vec{H})_z, \end{aligned} \tag{5}$$

where $\partial_z = \partial/\partial z$, $k_x$ and $k_y$ are the wavevector transverse components, $k_0 = \omega/c$ is the free-space wavenumber, $j = 1, 2$, $\hat{\varepsilon}_1$ and $\hat{\mu}_2$ are the tensors with $\varepsilon_1$ and $\mu_2$ on their main diagonal and zeros elsewhere, respectively ($\hat{\varepsilon}_1 = \varepsilon_1 \hat{I}$, $\hat{\mu}_2 = \mu_2 \hat{I}$, $\hat{I}$ is the identity tensor). From six components of the electromagnetic field $\vec{E}$ and $\vec{H}$, only four are independent. Thus the components $E_z$ and $H_z$ can be eliminated from the system (5) and derived a set of four first-order linear differential equations related to the transversal field components inside each layer of the structure [5]. For the ferrite ($0 \leq z \leq d_1$) and semiconductor ($d_1 \leq z \leq L$) layers these systems, respectively, are:

$$\partial_z \begin{pmatrix} E_x \\ E_y \\ H_x \\ H_y \end{pmatrix} = ik_0 \begin{pmatrix} 0 & 0 & k_x k_y / k_0^2 \varepsilon_1 + i\alpha & \mu_1^T - k_x^2 / k_0^2 \varepsilon_1 \\ 0 & 0 & -\mu_1^T + k_y^2 / k_0^2 \varepsilon_1 & -k_x k_y / k_0^2 \varepsilon_1 + i\alpha \\ -k_x k_y / k_0^2 \mu_1^L & -\varepsilon_1 k_0^2 + k_x^2 / k_0^2 \mu_1^L & 0 & 0 \\ \varepsilon_1 - k_y^2 / k_0^2 \mu_1^L & k_x k_y / k_0^2 \mu_1^L & 0 & 0 \end{pmatrix} \begin{pmatrix} E_x \\ E_y \\ H_x \\ H_y \end{pmatrix},$$
(6)

$$\partial_z \begin{pmatrix} E_x \\ E_y \\ H_x \\ H_y \end{pmatrix} = ik_0 \begin{pmatrix} 0 & 0 & k_x k_y / k_0^2 \varepsilon_2^L & \mu_2 - k_x^2 / k_0^2 \varepsilon_2^L \\ 0 & 0 & -\mu_2 + k_y^2 / k_0^2 \varepsilon_2^L & -k_x k_y / k_0^2 \varepsilon_2^L \\ -k_x k_y / k_0^2 \mu_2 - i\beta - \varepsilon_2^T + k_x^2 / k_0^2 \mu_2 & 0 & 0 \\ \varepsilon_2^T - k_y^2 / k_0^2 \mu_2 & k_x k_y / k_0^2 \mu_2 - i\beta & 0 & 0 \end{pmatrix} \begin{pmatrix} E_x \\ E_y \\ H_x \\ H_y \end{pmatrix}.$$
(7)

The sets of Eqs. (6) and (7) can be abbreviated by using a matrix formulation:

$$\partial_z \vec{\Phi}(z) = ik_0 \mathbf{A}(z) \vec{\Phi}(z), \quad 0 < z < L. \tag{8}$$

In this equation, $\vec{\Phi} = \{E_x, E_y, H_x, H_y\}^T$ is a four-component column vector (here upper index $T$ denotes the matrix transpose operator), while the 4×4 matrix function $\mathbf{A}(z)$ is piecewise uniform as

$$\mathbf{A}(z) = \begin{cases} \mathbf{A}_1, & 0 < z < d_1, \\ \mathbf{A}_2, & d_1 < z < L, \end{cases} \tag{9}$$

where the matrices $\mathbf{A}_1$ and $\mathbf{A}_2$ correspond to Eqs. (6) and (7), respectively.

Since the vector $\vec{\Phi}$ is known in the plane $z = 0$, the Eq. (7) is related to the Cauchy problem [25] whose solution is straightforward, because the matrix $\mathbf{A}(z)$ is piecewise uniform. Thus, the field components referred to boundaries of the double-layer period of the structure are related as[1]

$$\vec{\Phi}(L) = \mathbf{M}_2 \vec{\Phi}(d_1) = \mathbf{M}_2 \mathbf{M}_1 \vec{\Phi}(0) = \mathbb{M} \vec{\Phi}(0) = \exp[ik_0 \mathbf{A}_2 d_2] \exp[ik_0 \mathbf{A}_1 d_1] \vec{\Phi}(0), \quad (10)$$

where $\mathbf{M}_j$ and $\mathbb{M}$ are the transfer matrices of the corresponding layer and the period, respectively.

Suppose that $\gamma_j$ is the eigenvalue of the corresponding matrix $k_0 \mathbf{A}_j$ ($\det[k_0 \mathbf{A}_j - \gamma_j \mathbf{I}] = 0$), $j = 1, 2$ and $\mathbf{I}$ is the $4 \times 4$ identity matrix. When $|\gamma_j| d_j \ll 1$ (i.e., both layers in the period are electrically thin), the next long-wave approximations can be used [26]

$$\exp[ik_0 \mathbf{A}_2 d_2] \exp[ik_0 \mathbf{A}_1 d_1] \simeq \mathbf{I} + ik_0 \mathbf{A}_1 d_1 + ik_0 \mathbf{A}_2 d_2. \quad (11)$$

Let us now consider a single layer of effective permittivity $\hat{\varepsilon}_e$, effective permeability $\hat{\mu}_e$ and thickness $L$. Quantity $\mathbf{A}_e$ is defined in a way similar to (6), (7):

$$\partial_z \begin{pmatrix} E_x \\ E_y \\ H_x \\ H_y \end{pmatrix} = ik_0 \begin{pmatrix} 0 & 0 \\ 0 & 0 \\ -k_x k_y / k_0^2 \mu_e^L - i\beta_e & -\varepsilon_e^T + k_x^2 / \mu_e^L k_0^2 \\ \varepsilon_e^T - k_y^2 / k_0^2 \mu_e^L & k_x k_y / k_0^2 \mu_e^L - i\beta_e \\ & k_x k_y / k_0^2 \varepsilon_e^L + i\alpha_e & \mu_e^T - k_x^2 / k_0^2 \varepsilon_e^L \\ & -\mu_e^T + k_y^2 / k_0^2 \varepsilon_e^L & -k_x k_y / k_0^2 \varepsilon_e^L + i\alpha_e \\ & 0 & 0 \\ & 0 & 0 \end{pmatrix} \begin{pmatrix} E_x \\ E_y \\ H_x \\ H_y \end{pmatrix}, \quad (12)$$

and (10):

$$\vec{\Phi}(L) = \mathbf{M}_e \vec{\Phi}(0) = \mathbb{M} \vec{\Phi}(0) = \exp[ik_0 \mathbf{A}_e L] \vec{\Phi}(0). \quad (13)$$

---

[1] The series $\exp(\mathbf{X}) = \mathbf{I} + \sum_{m=1}^{\infty} \frac{1}{m!} \mathbf{X}^m$ converges for square matrices $\mathbf{X}$, i.e. function $\exp(\mathbf{X})$ is defined for all square matrices [25].

Provided that $\gamma_e$ is the eigenvalue of the matrix $k_0 \mathbf{A}_e$ ($\det[k_0\mathbf{A}_e - \gamma_e \mathbf{I}] = 0$) and $|\gamma_e|L \ll 1$ (i.e., the entire composite layer is electrically thin as well), the next approximation follows

$$\exp[ik_0\mathbf{A}_e L] \simeq \mathbf{I} + ik_0\mathbf{A}_e L. \qquad (14)$$

Equations (11) and (14) permit us to establish the following equivalence between bilayer and single layer:

$$\mathbf{A}_e = f_1\mathbf{A}_1 + f_2\mathbf{A}_2, \quad f_j = d_j/L. \qquad (15)$$

In the case when the directions of both wave propagation and static magnetic field are coincident ($k_x = k_y = 0$), the following simple expressions for the effective constitutive parameters of the homogenized medium can be obtained:

$$\mu_e^T = f_1\mu_1^T + f_2\mu_2, \quad \varepsilon_e^T = f_1\varepsilon_1 + f_2\varepsilon_2^T, \quad \alpha_e = f_1\alpha, \quad \beta_e = f_2\beta. \qquad (16)$$

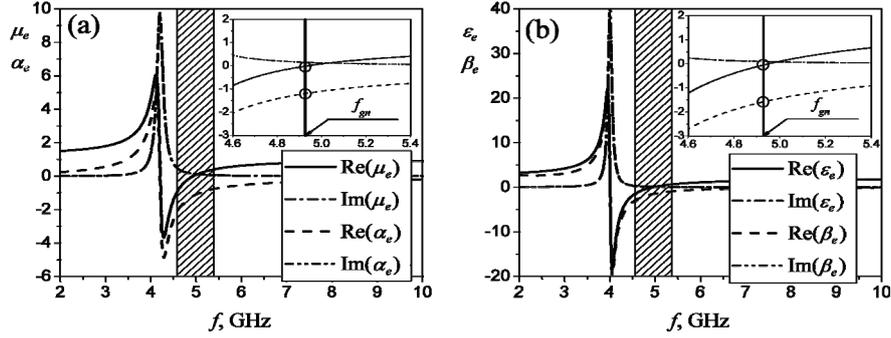

**Fig. 3.** Frequency dependences of (**a**) effective permeability and (**b**) effective permittivity of the homogenized ferrite-semiconductor medium. Parameters of the ferrite and semiconductor layers are the same as in Fig. 2; $d_1$ = 0.05 mm, $d_2$ = 0.2 mm. The circles mark the situation when $\mathrm{Re}(\mu_e^T)$ and $\mathrm{Re}(\varepsilon_e^T)$ are close to zero while $\mathrm{Re}(\alpha_e) \neq 0$, $\mathrm{Re}(\beta_e) \neq 0$ and losses in the ferrite and semiconductor layers are small.

The effective constitutive parameters calculated according to the formula (16) are given in Fig. 3. The whole frequency range can be divided into three specific bands where parameters of the tensors $\hat{\mu}_e$ and $\hat{\varepsilon}_e$ acquire different properties. In the first band, located between 2 GHz and 3 GHz, $\mu_e^T$, $\varepsilon_e^T$, $\alpha_e$ and $\beta_e$ have positive values of their real parts and small imaginary parts. In the second band, between 3 GHz and 4.5 GHz, the real parts of parameters vary from positive values to negative ones as the frequency increases. These transitions occur at the frequencies of the ferromagnetic resonance of ferrite ($f_{fr}$ = 4.2 GHz) and the cyclo-tron resonance of semiconductor $f_{pr}$ = 4.0 GHz), respectively. In this band the medium losses are very significant. Finally, in the third frequency band, located from 4.5 GHz to 5.5 GHz, the real parts of parameters have a transition from negative to positive values while their imaginary parts are small. The latter band is given in the insets of Fig. 3 on a larger scale. One can see that there is a frequency $f_{gn} \approx 4.94$ GHz where real parts of $\mu_e^T$ and $\varepsilon_e^T$ simultaneously reach zero. It is significant that, by special adjusting ferrite and semiconductor type, external static magnetic field strength and thicknesses of layers, it is possible to obtain the condition when real parts of $\mu_e^T$ and $\varepsilon_e^T$ acquire zero at the same frequency. Exactly this situation is marked in the insets of Fig. 3 with circles. Note that at this frequency, the real parts of $\alpha_e$ and $\beta_e$ are far from zero and the medium losses are small.

### *1.2.4 Eigenvalue Problem*

The formulation of the eigenvalue problem on the matrix $\mathbf{A}_e$ ( $\det[\mathbf{A}_e - \eta_e \mathbf{I}] = 0$ ), whose coefficients are defined as (16), gives us the characteristic equation on the effective refractive index $\eta_e$ of the medium:

$$\eta_e^4 - 2\eta_e^2 \left( \varepsilon_e^T \mu_e^T + \alpha_e \beta_e \right) + \left( \varepsilon_e^T \mu_e^T \right)^2 - \left( \mu_e^T \beta_e \right)^2 - \left( \varepsilon_e^T \alpha_e \right)^2 + \left( \alpha_e \beta_e \right)^2 = 0 , \quad (17)$$

whose solutions are

$$(\eta_e^\pm)^2 = \left( \varepsilon_e^T \pm \beta_e \right)\left( \mu_e^T \pm \alpha_e \right) = \varepsilon^\pm \mu^\pm . \tag{18}$$

Here the signs '$\pm$' are related to two different eigenwaves with propagation constants $\gamma_e^\pm = k_0 \eta_e^\pm$. It is well known that in an unbounded gyrotropic medium they have right (RCP, $\gamma_e^+$) and left (LCP, $\gamma_e^-$) circular polarizations [21].

Especially interesting situation appears if the real parts of $\mu_e^T$ and $\varepsilon_e^T$ are close to zero and the medium losses are small. In this case there is $|\varepsilon^\pm \mu^\pm| \approx |\alpha_e' \beta_e'|$, and the propagation constants become as:

$$\gamma_e = -\gamma_e^+ = \gamma_e^- \approx k_0 \sqrt{|\alpha_e' \beta_e'|} . \tag{19}$$

Thus, the propagation constants of the RCP ($\gamma_e^+$) and LCP ($\gamma_e^-$) waves are equal in the magnitude but opposite in sign to each other, and the *backward propagation* appears for the RCP wave while for the LCP wave it is forward one (see also,

Ch. IV of [27]). Recall that the backward wave is the wave in which the direction of the Poynting vector is opposite to that of its phase velocity [28]. The similar peculiarity of the RCP and LCP waves propagation occurs also in the chiral-nihility media [2]-[4], so in the analogy with them, the condition (19) is related to the *gyrotropic-nihility* media [11]. The frequency band, at which the gyrotropic-nihility condition is satisfied for the RCP wave, is depicted in Fig. 4. Particularly, the gyrotropic-nihility frequency $f_{gn}$ is marked in the inset of this figure with circles.

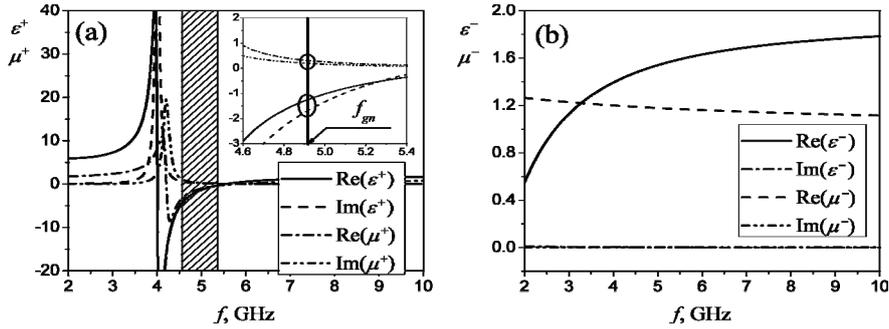

**Fig. 4.** Frequency dependences of the material parameters of the equivalent gyrotropic medium for the (**a**) RCP and (**b**) LCP eigenwaves. Parameters of the ferrite and semiconductor layers are the same as in Fig. 2; $d_1 = 0.05$ mm, $d_2 = 0.2$ mm.

## 1.3 Numerical Results: Reflected and Transmitted Fields

### 1.3.1 Spectral and Angular Behaviors

It is anticipated that if the frequency of the electromagnetic wave which incidents on a finite layer of such a composite medium is chosen to be nearly the frequency $f_{gn}$ of the gyrotropic-nihility condition, the transmitted and reflected fields will acquire some unusual properties. In order to demonstrate this, in the long-wavelength limit, the reflection and transmission coefficients can equivalently be calculated using the rigorous solution (10) or the approximate solution (13) of Eq. (8) because these solutions give the same result. In particular, we are interested here in the study of the energy relations between the transmitted and reflected fields, i.e. the polarization effects are not discussed in this chapter. Nevertheless we refer the reader to Ref. [11] where some polarization features of the studied structure are revealed.

So, the transmittance is calculated as a function of the frequency and the angle of incidence (Fig. 5a). One can see that this function exhibits an expanded flat area of the transmittance at the frequency where the gyrotropic-nihility condition holds. At this frequency the complete transmission of the partial plane monochromatic waves takes a place almost in the entire range of angles of incidence except the range of glancing angles. Such a high transmittance appears due to the peculiarities of the medium impedances ($Z^{\pm} = \sqrt{\mu^{\pm}/\varepsilon^{\pm}}$) related to the RCP and LCP waves. It is particularly remarkable that in the vicinity of the gyrotropic-nihility frequency $f_{gn}$, the parameters $\alpha_e$ and $\beta_e$ are close in value to each other and their real parts approach to unit which can be clearly seen in Fig. 3. It leads to the fact that the medium becomes to be impedance matched to free space [11]. Directly at the gyrotropic-nihility frequency, the impedances related to the RCP and LCP waves become indistinguishable: $Z = Z^{+} = Z^{-} = \sqrt{|\alpha'_e|/|\beta'_e|}$.

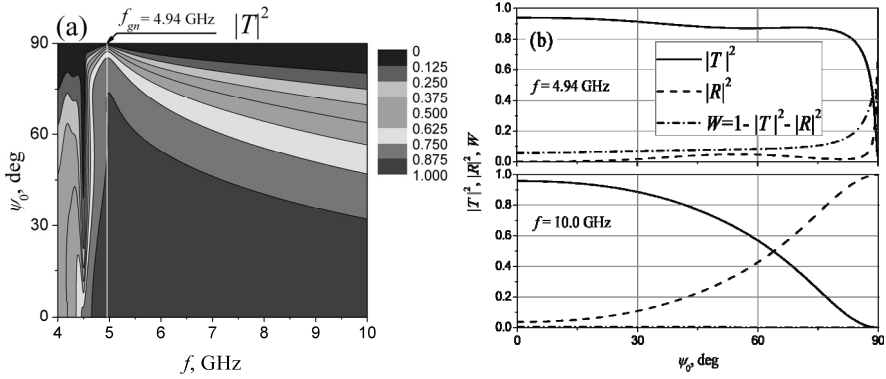

**Fig. 5.** (**a**) Transmittance as a function of the frequency and the angle of incidence of the plane monochromatic wave for the equivalent gyrotropic layer with finite thickness ($NL$ = 2.5 mm). (**b**) The angular dependences of the transmittance, reflectance and absorption coefficient at the gyrotropic-nihility frequency (top figure) and far from it (bottom figure). Parameters of the ferrite and semiconductor layers are the same as in Fig. 2; $d_1$ = 0.05 mm, $d_2$ = 0.2 mm.

This feature is also confirmed by the data plotted in Fig. 5b where the reflectance, transmittance and absorption coefficient are calculated at two different frequencies for a comparison. Thus the first frequency is chosen at the gyrotropic-nihility condition and the second one is selected to be far from the frequencies of gyrotropic-nihility condition and the ferromagnetic and plasma resonances. At the frequency of $f$ = 10 GHz, the curves have typical form where the transmittance monotonically decreases and the reflectance monotonically increases as the angle of incidence rises. On the other hand, at the gyrotropic-nihility frequency, the curves of the transmittance and reflectance are different drastically from that ones in the first case. Thus, the level of the transmittance/reflectance remains to be invariable almost down to the glancing angles. At the same time, the reflectance is small down to the glancing angles because the medium is impedance matched to free space.

## *1.3.2 Gaussian Beam Transmission*

Since the beam field is represented by an angular continuous spectrum of plane waves, the transmitted beam distribution depends on the angular characteristic of the transmission coefficient of spatial plane monochromatic waves at a particular frequency. We consider an incident Gaussian beam with the spectral density assigned due to the law $U(\vec{\kappa}_{in}) = \exp[-(\vec{w}\cdot\vec{\kappa}_{in})^2/16]H_m(k_{xin}w_x/\sqrt{2})H_n(k_{yin}w_y/\sqrt{2})$, where $\vec{w} = \{w_x, w_y\}$, $w_x$ and $w_y$ are the beam widths along $x_{in}$ and $y_{in}$ axis, respectively, $H_\upsilon(\cdot)$ is the Hermit polynomial of $\upsilon$-th order ($\upsilon = m, n$). In this chapter we restrict ourselves to the case of the zero-order ($m = n = 0$) beam. The final distribution of the transmitted beam is presented in Figs. 6, 7 in two- and three-dimensions. As before the results are obtained for two distinct frequencies.

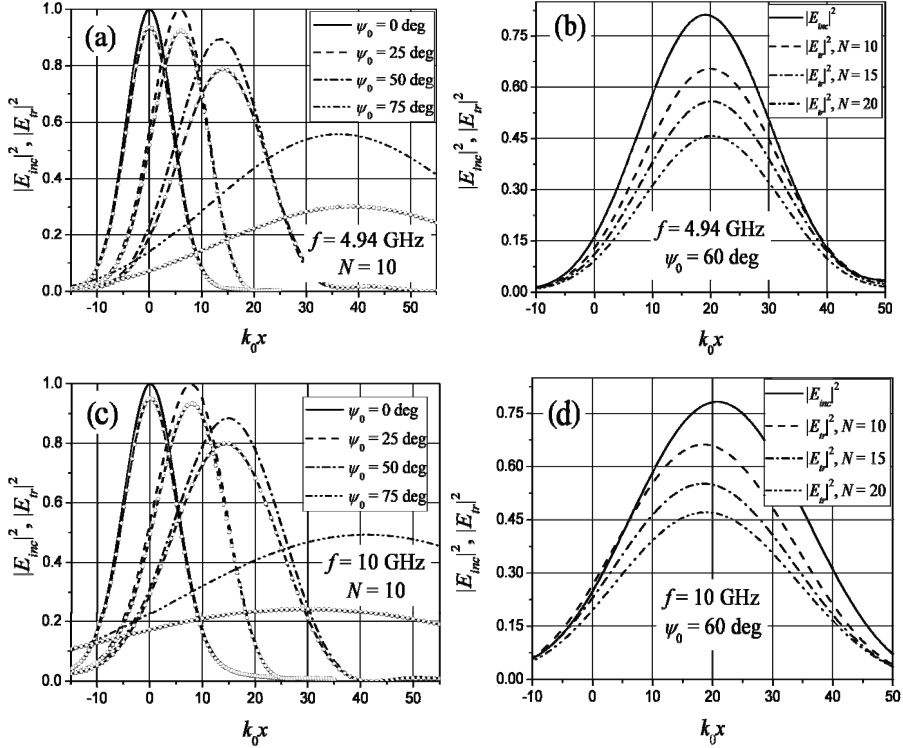

**Fig. 6.** The two-dimensional distribution of the absolute value of the incident beam $|E_{inc}|^2$ and the transmitted beam $|E_{tr}|^2$ for different (**a**, **c**) angles of incidence of the primary beam and (**b**, **d**) number of structure's periods. The field distribution is normalized to the maximum value of the normally incident beam. Parameters of the ferrite and semiconductor layers are the same as in Fig. 2. The incident beam parameters are: $k_0w_x = k_0h = 10$, $\varphi_{in} = 0$ deg. Other structure parameters are: $d_1 = 0.05$ mm, $d_2 = 0.2$ mm.

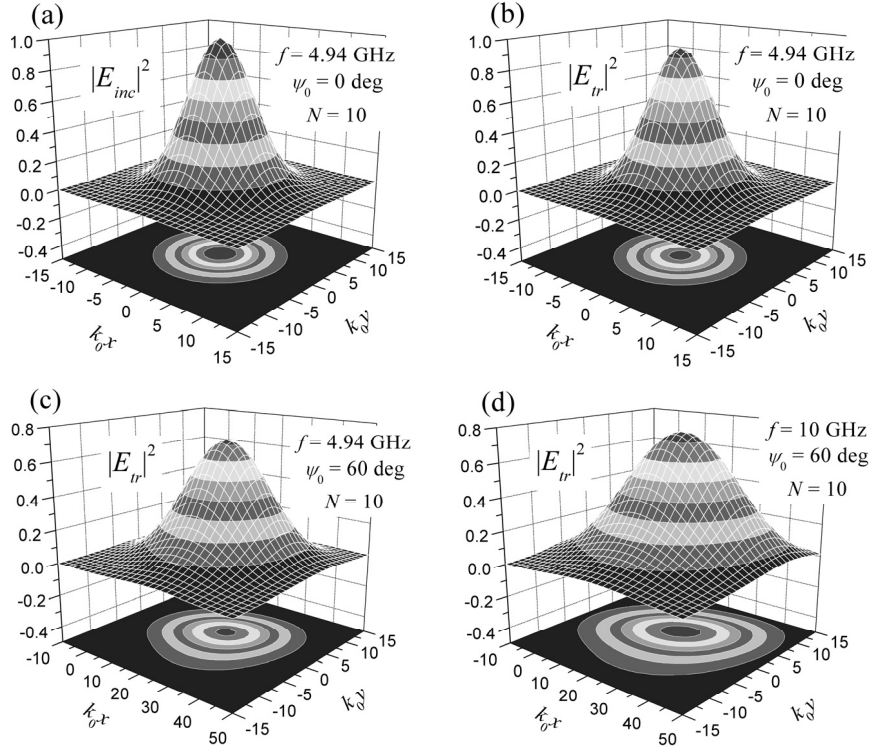

**Fig. 7.** The three-dimensional distribution of the absolute value of (**a**) the incident beam $|E_{inc}|^2$ and (**b-d**) the transmitted beam $|E_{tr}|^2$. The field distribution is normalized to the maximum value of the normally incident beam. Parameters of the ferrite and semiconductor layers are the same as in Fig. 2. The incident beam parameters are: $k_0w_x = k_0w_y = k_0h = 10$, $\varphi_{in} = 0$ deg. Other structure parameters are: $d_1 = 0.05$ mm, $d_2 = 0.2$ mm.

So, at the frequency $f_{gn}$ where the gyrotropic-nihility condition holds, the studied gyroelectromagnetic medium is well impedance matched to free space, and spatial plane monochromatic waves can completely pass through the system up to the glancing angles. As a result, the transmitted beam pattern does not acquire any significant distortion of its form nearly the frequency $f_{gn}$, while this feature is not inherent to the pattern of the transmitted beam to be far from this frequency. It should be noted that this shape retention of the transmitted beam pattern remains unchanged even under the oblique incidence of the primary beam and this effect is polarization insensitive.

In conclusion, the peculiarities of the Gaussian beam interaction with a ferrite-semiconductor finely-stratified structure being under an action of an external static magnetic field in the Faraday geometry is presented in this chapter. In the long-wavelength limit, when the structure layers are optically thin, the effective medium theory is developed, and the effective constitutive parameters of the equiva-lent uniform anisotropic medium are obtained analytically. On the basis of these parameters the peculiarities of the eigenwaves propagation are studied and the possibility of achieving the gyrotropic-nihility condition is predicted.

The reflection, transmission and absorption of waves in the system are studied in vicinity of the gyrotropic-nihility frequency. It is found out that under the oblique incidence of the spatial plane monochromatic wave on the studied structure, the level of the transmission/reflection remains to be invariable almost down to the glancing angles when the gyrotropic-nihility condition is satisfied. As a result, at the frequency of the gyrotropic-nihility condition the Gaussian beam can pass through such a system keeping its parameters unchanged (beam width and shape) except of a portion of the absorbed energy even under the oblique incidence of the primary beam.

**Acknowledgments** This work was partially supported (V.R. Tuz) by Ministry of Education and Science of Ukraine under the Program "Electrodynamics of layered composites with chiral properties and multifunctional planar systems", Project No. 0112 U 000561.

# References

1. A. Lakhtakia, An electromagnetic trinity from "negative permittivity" and "negative permeability", *Int. J. Infrared Millimeter Waves,* **22**, 1731-1734, (2001)
1. S. Tretyakov, I. Nefedov, A.H. Sihvola, S. Maslovki, C. Simovski, Waves and energy in chiral nihility, *J. Electromagn. Waves Appl.*, **17**, 695-706, (2003)
2. C.-W. Qiu, N. Burokur, S. Zouhdi, L.-W. Li, Chiral nihility effects on energy flow in chiral materials, *J. Opt. Soc. Am. A,* **25**, 55-63, (2008)
3. V. Tuz, C.-W. Qiu, Semi-infnite chiral nihility photonics: Parametric dependence, wave tunelling and rejection, *Prog. Electromagn. Res.*, **103**, 139-152, (2010)
4. V.R. Tuz, M.Y. Vidil, S.L. Prosvirnin, Polarization transformations by a magneto-photonic layered structure in the vicinity of a ferromagnetic resonance, *J. Opt.,* **12**, 095102, (2010)
5. E. Prati, Propagation in gyroelectromagnetic guiding systems, *J. Electromagn. Waves Appl.*, **17**, 1177-1196, (2003)
6. R.H. Tarkhanyan, D.G. Niarchos, Effective negative refractive index in ferromagnet-semiconductor superlattices, *Opt. Express*, **14**, 5433-5444, (2006)
7. A.V. Ivanova, O.A. Kotelnikova, V.A. Ivanov, Gyrotropic left-handed media: Energy flux and circular dichroism, *J. Magn. Magn. Mat.*, **300**, e67-e69, (2006)
8. R.X. Wu, T. Zhao, J.Q. Xiao, Periodic ferrite-semiconductor layered composite with negative index of refraction," *J. Phys.: Condens. Matter.*, **19**, 026211, (2007)
9. O.V. Shramkova, Transmission spectra in ferrite-semiconductor periodic structure, *Prog. Electromagn. Res. M*, **7**, 71-85, (2009)
10. V.R. Tuz, O.D. Batrakov, Y. Zheng, Gyrotropic-nihility in ferrite-semi-conductor composite in Faraday geometry, *Prog. Electromagn. Res. B*, **41**, 397-417, (2012)
11. Yi Jin, Sailing He, Focusing by a slab of chiral medium, *Opt. Express*, **13**, 4974-4979, (2005)
12. S.N. Shulga, Two-dimensional wave beam scattering on an anisotropic half-space with anisotropic inclusion, *Optics and Spectroscopy*, **87**, 503-509, (1999)

13. A.V. Malyuskin, D.N. Goryushko, S.N. Shulga, A.A. Shmatko, Scattering of a wave beam by inhomogeneous anisotropic chiral layer, *Int. Conf. Math. Methods EM Theory (MMET 2002), Sep. 10-13, Kyiv, Ukraine,* 566-568, (2002)
14. T.M. Grzegorczyk, X. Chen, J. Pacheco Jr., J. Chen, B.-I. Wu, J.A. Kong, Reflection coefficients and Goos-Hänchen shifts in anisotropic and bianisotropic left-handed metamaterials, *Prog. Electromagn. Res.*, **51**, 83–113, (2005)
15. W.T. Dong, L. Gao, C.-W. Qiu, Goos-Hänchen shift at the surface of chiral negative refractive media, *Prog. Electromagn. Res.*, **90**, 255-268 (2009)
16. R.-L. Chern, P.-H. Chang, Negative refraction and backward wave in pseudochiral mediums: Illustrations of Gaussian beams, *Opt. Express*, **21**, 2657-2666, (2013)
17. R.-L. Chern, P.-H. Chang, Negative refraction and backward wave in chiral mediums: Illustrations of Gaussian beams, *J. Appl. Phys.*, **113**, 153504, (2013)
18. K.M. Luk, A.L. Cullen, Three-dimensional Gaussian beam reflection from short-circuited isotropic ferrite slab, *IEEE. Trans. Antennas Propag.,* **41**, 962-966, (1993)
19. V. Tuz, Three-dimensional Gaussian beam scattering from a periodic sequence of bi-isotropic and material layers, *Prog. Electromagn. Res.* B, **7**, 53-73, (2008)
20. A.G. Gurevich, *Ferrites at Microwave Frequencies* (Heywood, 1963)
21. R.E. Collin, *Foundation for Microwave Engineering* (Wiley, 1992)
22. F.G. Bass, A.A. Bulgakov, *Kinetic and Electrodynamic Phenomena in Classical and Quantum Semiconductor Superlattices* (Nova Science, 1997)
23. G.A. Korn, T.M. Korn, *Mathematical Handbook for Scientists and Engineers: Definitions, Theorems, and Formulas for Reference and Review* (McGraw-Hill Book Co., 1968)
24. V.A. Jakubovich, V.H. Starzhinskij, *Linear Differential Equations with Periodic Coefficients* (Wiley, New York, 1975)
25. A. Lakhtakia, C.M. Krowne, Restricted equivalence of paired epsilon-negative and mu-negative layers to a negative phase-velocity materials (alias left-handed material), *Optik*, **114**, 305–307, (2003)
26. V.G. Veselago, The electrodynamics of substances with simultaneously negative values of $\varepsilon$ and $\mu$, *Sov. Phys. Usp.*, **10**, 509–514, (1968)
27. V.V. Shevchenko, Forward and backward waves: Three definitions and their interrelation and applicability, *Phys.-Usp.*, **50**, 287–290, (2007)